\documentclass[12pt]{article}
\usepackage{epsf,latexsym}
\usepackage{amsmath,amssymb,amsthm,cite}
\usepackage{graphicx}

\theoremstyle{definition}                    

\theoremstyle{remark}

\numberwithin{equation}{section}             

\newcommand{\de}{\hbox{\rm{d}}}

\newcommand{\bb}{\begin{eqnarray}}
\newcommand{\ee}{\end{eqnarray}}
\newcommand{\eee}{\nonumber\end{eqnarray}}
\newcommand{\qq}{\quad}

\usepackage[ps,dvips,arc,frame]{xy}

\epsfverbosetrue
\newcommand{\ddf}{\hbox{$^f$\hspace{-0.15cm} $\mathcal{D}$}}

\newcommand{\T}{{\rm tr}}

\newcommand{\pp}[1]{\begin{pmatrix} #1 \end{pmatrix}}

\newcommand{\rxyh}[2]{{\begin{xy} 0;<2mm,0mm>:<0mm,2mm>::0;0,
,(5,-2)*{a}
,(10,-2)*{b}
,(15,-1.8)*{\bar{b}}
,(20,-2)*{c}
,(25,-1.8)*{d}
,(30,-1.8)*{\bar{d}}
,(2,-5)*{a}
,(2,-10)*{b}
,(1.8,-15)*{\bar{b}}
,(2,-20)*{c}
,(1.8,-25)*{d}
,(1.8,-30)*{\bar{d}}
,(5,-5)*\cir(#1,0){}
,(10,-5)*\cir(#1,0){}
,(15,-5)*\cir(#1,0){}
,(20,-5)*\cir(#1,0){}
,(25,-5)*\cir(#1,0){}
,(30,-5)*\cir(#1,0){}
,(5,-10)*\cir(#1,0){}
,(10,-10)*\cir(#1,0){}
,(15,-10)*\cir(#1,0){}
,(20,-10)*\cir(#1,0){}
,(25,-10)*\cir(#1,0){}
,(30,-10)*\cir(#1,0){}
,(5,-15)*\cir(#1,0){}
,(10,-15)*\cir(#1,0){}
,(15,-15)*\cir(#1,0){}
,(20,-15)*\cir(#1,0){}
,(25,-15)*\cir(#1,0){}
,(30,-15)*\cir(#1,0){}
,(5,-20)*\cir(#1,0){}
,(10,-20)*\cir(#1,0){}
,(15,-20)*\cir(#1,0){}
,(20,-20)*\cir(#1,0){}
,(25,-20)*\cir(#1,0){}
,(30,-20)*\cir(#1,0){}
,(5,-25)*\cir(#1,0){}
,(10,-25)*\cir(#1,0){}
,(15,-25)*\cir(#1,0){}
,(20,-25)*\cir(#1,0){}
,(25,-25)*\cir(#1,0){}
,(30,-25)*\cir(#1,0){}
,(5,-30)*\cir(#1,0){}
,(10,-30)*\cir(#1,0){}
,(15,-30)*\cir(#1,0){}
,(20,-30)*\cir(#1,0){}
,(25,-30)*\cir(#1,0){}
,(30,-30)*\cir(#1,0){}
#2\end{xy}}}

\begin{document}

\thispagestyle{empty}

\begin{center}
CENTRE DE PHYSIQUE TH\'EORIQUE \footnote{\, Unit\'e Mixed de
Recherche (UMR) 6207 du CNRS et des Universit\'es Aix-Marseille 1 et 2 \\ \indent \quad \, Sud Toulon-Var, Laboratoire affili\'e \`a la 
FRUMAM (FR 2291)} \\ CNRS--Luminy, Case 907\\ 13288 Marseille Cedex 9\\
FRANCE
\end{center}

\vspace{1.5cm}

\begin{center}
{\Large\textbf{Almost-Commutative Geometries \\ Beyond the Standard Model III:
\\[2mm] Vector Doublets}} 
\end{center}

\vspace{1.5cm}

\begin{center}
{\large Romain Squellari$^1$, Christoph A. Stephan$^2$ }

\vspace{1.5cm}

{\large\textbf{Abstract}}
\end{center}

We will present a new extension of the standard model of particle physics 
in its almost-commutative formulation. This extension has as its basis
the algebra of the standard model with four summands \cite{class}, 
and enlarges only the particle content by an arbitrary number of generations
of left-right symmetric doublets which couple vectorially to the 
$U(1)_Y\times SU(2)_w$ subgroup of the standard model.

As in the model presented in \cite{theta}, which introduced particles with a new 
colour,  grand unification is no longer required by the  spectral action. The new
model may also possess a candidate for dark matter  in the hundred 
TeV mass range with neutrino-like cross section.

\vspace{0.5cm}

\vskip 1truecm
\indent
CPT-P28-2007 \\
PACS-92: 11.15 Gauge field theories\\
MSC-91: 81T13 Yang-Mills and other gauge theories

\vskip 1truecm

\vspace{1cm}
\noindent 
$^1$ also at Universit\'e Aix--Marseille 1,
rsquella@cpt.univ-mrs.fr \\
$^2$ also at Universit\'e Aix--Marseille 1,
christophstephan@gmx.de\\

\newpage

\section{Introduction}

In this paper we present an extension of the standard model
by an arbitrary number of left-right symmetric doublets which couple vectorially to the 
$U(1)_Y\times SU(2)_w$ subgroup of the standard model. 
We will call these particles for simplicity {\it vector doublets}. This extension
is done within the framework of noncommutative geometry
which was pioneered by Alain Connes \cite{book}. 
It extends the standard model as presented in the recent
formulation of almost-commutative geometry \cite{c06,barr,mc2}
which is a slight modification of the original formulation \cite{cc}.

The model treated here is the third viable extension of the 
standard model, following the $AC$ model \cite{chris} and
a model which realises Okun's $\theta$-particles \cite{okun}
within almost-commutative geometry \cite{theta}. These
extensions are very rare because the constraints on the models
from the axioms of almost commutative geometry and the
spectral action are severe. Nevertheless at least the
$AC$ model exhibits a viable dark matter candidate \cite{klop}.
This might also be true for the vector doublet model presented here.

As a basis for the vector doublet model we take the formulation
of the the standard model with four summands in the matrix
algebra which was found in the classification of almost-commutative
geometries \cite{class,prog}. The vector doublet model has many
similarities to the $\theta$-model, notably the constraints on 
the gauge couplings of 
the model no longer resemble those
of grand unified theories.  

Adding vector doublets has a rather small  effect on the Higgs mass 
but lowers the cut-off scale of the spectral action considerably. 
This may provide a natural explanation for the possible masses
of the vector doublets which are in the upper TeV scale.
Here we will not consider mixing of the generations, but this should certainly
be investigated more closely since it may provide a clue to 
the matter-antimatter asymmetry of the universe.

The paper is organised as follows: We first give the basic notions
of a spectral triple, the main building  block of noncommutative geometry.
Then we quickly review how the Yang-Mills-Higgs model is obtained
via the fluctuated Dirac operator and the spectral action. This 
account is far from exhaustive and we refer to \cite{cc,farewell,mc2} for
a detailed presentation.

For the vector doublets the details of the spectral triple and the lift of 
the automorphisms are given.The Lagrangian as well as the 
constraints on the couplings are calculated and we give
a short summary of the mass splitting of the doublet components
due to radiative corrections. With help of the
one-loop renormalisation group equations  the Higgs boson mass and
the cut-off scales are calculated for up to three generations of
vector doublets.
\section{Finite Spectral Triples}

In this section we will give the necessary basic definitions  of almost commutative
geometries from a particle physics point of view. 
For our calculations, only the finite part matters, 
so we restrict ourselves to real, finite spectral triples in $KO$-dimension
six:
($\mathcal{A},\mathcal{H},\mathcal{D}, $ $J,\chi$). Note that in the literature
before \cite{c06,barr,mc2} the finite part of the spectral triple was
considered to be of $KO$-dimension zero. The change in this algebraic dimension
amounts in some  sign changes, i.e. the commutator for the real structure
and the chirality changes into an anti-commutator and the anti-particles 
have opposite chirality with respect to the particles.

\subsection{Basic Definitions}

The algebra $\mathcal{A}$ is
a finite sum of matrix algebras
$\mathcal{A}= \oplus_{i=1}^{N} M_{n_i}(\mathbb{K}_i)$ with $\mathbb{K}_i=\mathbb{R},\mathbb{C},\mathbb{H}$ where $\mathbb{H}$
denotes the quaternions. 
A faithful representation $\rho$ of $\mathcal{A}$ is given on the finite dimensional Hilbert space $\mathcal{H}$.
The Dirac operator $\mathcal{D}$ is a selfadjoint operator on $\mathcal{H}$ and plays the role of the fermionic mass matrix.
$J$ is an antiunitary involution, $J^2=1$, and is interpreted as the charge conjugation
operator of particle physics.
The chirality $\chi$  is a unitary involution, $\chi^2=1$, whose eigenstates with eigenvalue
$+1$ $(-1)$ are interpreted as right (left) particle states and  $-1$ $(+1)$ for 
right (left) antiparticle states.
These operators are required to fulfill Connes' axioms for spectral triples:

\begin{itemize}
\item  $[J,\mathcal{D}]=\{J,\chi\}=\{ \mathcal{D},\chi\} =0$, 
 
$[\chi,\rho(a)]=[\rho(a),J\rho(b)J^{-1}]=
[[\mathcal{D},\rho(a)],J\rho(b)J^{-1}]=0, \forall a,b \in \mathcal{A}$.
\item The chirality can be written as a finite sum $\chi =\sum_i\rho(a_i)J\rho(b_i)J^{-1}.$
This condition is called {\it orientability}.
\item The intersection form
$\cap_{ij}:=\T(\chi \,\rho (p_i) J \rho (p_j) J^{-1})$ is non-degenerate,
$\rm{det}\,\cap\not=0$. The
$p_i$ are minimal rank projections in $\mathcal{A}$. This condition is called
{\it Poincar\'e duality}.
\end{itemize} 
Now the Hilbert space $\mathcal{H}$ and the representation $\rho$ are  decomposed 
into left and right, particle and antiparticle spinors and representations:
\begin{eqnarray}
\mathcal{H}=\mathcal{H}_L\oplus\mathcal{H}_R\oplus\mathcal{H}_L^c\oplus\mathcal{H}_R^c \quad \quad 
\rho = \rho_L \oplus \rho_R \oplus \overline{ \rho_L^c} \oplus \overline{ \rho_R^c}
\notag
\label{representation}
\end{eqnarray}
In this representation the Dirac operator has the form
\begin{eqnarray}
\mathcal{D}=\pp{0&\mathcal{M}&0&0\\
\mathcal{M}^*&0&0&0\\ 0&0&0&\overline{\mathcal{M}}\\
0&0&\overline{\mathcal{M}^*}&0}, \label{opdirac}
\notag
\end{eqnarray}
where $\mathcal{M}$ is the fermionic mass matrix connecting the left and the right handed fermions.

Since the individual matrix algebras have only one fundamental representation for $\mathbb{K}=
\mathbb{R},\mathbb{H}$ and two for $\mathbb{K}=\mathbb{C}$ (the fundamental one and its complex
conjugate), $\rho$ may be written as a direct sum of these fundamental representations with
mulitiplicities
\begin{eqnarray}
\rho(\oplus_{i=1}^N a_i):=(\oplus_{i,j=1}^N
a_i \otimes
1_{m_{ji}} \otimes 1_{(n_j)})\
\oplus\ ( \oplus_{i,j=1}^N 1_{(n_i)} \otimes 1_{m_{ji}} \otimes
\overline{a_j} ).
\nonumber
\end{eqnarray}
The first summand denotes the particle sector and the second the antiparticle sector. For the dimensions
of the unity matrices we have $(n)=n$ for $\mathbb{K}=\mathbb{R},\mathbb{C}$ and $(n)=2n$ for
$\mathbb{K}=\mathbb{H}$ and the convention $1_0=0$.
The multiplicities $m_{ji}$ are non-negative integers. Acting with the real structure
$J$ on $\rho$ permutes the main summands and complex conjugates them. It is also possible to write
the chirality as a direct sum
\begin{eqnarray}
\chi=(\oplus_{i,j=1}^N 1_{(n_i)} \otimes \chi_{ji}1_{m_{ji}} \otimes
1_{(n_j)})\  
\oplus\ (\oplus_{i,j=1}^N 1_{(n_i)} \otimes (-\chi_{ji})1_{m_{ji}} \otimes 1_{(n_j)}),
\nonumber
\end{eqnarray}
where $\chi_{ji}=\pm 1$ according to the previous convention on left-(right-)handed spinors.
One can now define the multiplicity matrix $\mu \in M_N(\mathbb{Z})$ such that
$\mu _{ji}:=\chi _ {ji}\, m_{ji}$. This matrix is symmetric and decomposes into a particle and an antiparticle matrix, the second being just the particle matrix transposed, $\mu= \mu_P + \mu_A = \mu_P - \mu_P^T$. The intersection form of the Poincar\'e duality is now simply $\cap = \mu - \mu^T$, see \cite{kps}. Note that in contrast to
the case of $KO$-dimension zero, the multiplicity matrix is now antisymmetric rather
than symmetric.

\subsection{Obtaining the Yang-Mills-Higgs theory}

To complete our short survey on the almost-commutative standard model, we will give a brief glimpse
on how to construct the actual Yang-Mills-Higgs theory.
We started out with the fixed (for convenience flat) Dirac operator of a 4-dimensional spacetime with a fixed fermionic
mass matrix. To generate curvature we have to perform a general coordinate transformation and then
fluctuate the Dirac operator. This can be achieved by lifting the automorphisms of the algebra to
the Hilbert space, unitarily transforming the Dirac operator with the lifted automorphisms
and then building linear combinations. Again we restrict ourselves to the finite case.
Except for complex conjugation in $M_n(\mathbb{C})$ and permutations of
identical summands in the algebra $\mathcal{A}=\mathcal{A}_1\oplus\mathcal{A}_2\oplus ...\oplus\mathcal{A}_N$,
every algebra automorphism
$\sigma
$  is inner, $\sigma (a)=uau^{-1}$ for a unitary $ u\in U(\mathcal{A})$. Therefore
the connected component of the automorphism group is
Aut$(\mathcal{A})^e=U(\mathcal{A})/(U(\mathcal{A})\cap{\rm Center}(\mathcal{A}))$. Its lift to the Hilbert
space \cite{real}
\bb 
\mathbb{L}(\sigma )=\rho (u)J\rho (u)J^{-1}
\eee is multi-valued. To avoid the multi-valuedness in the fluctuations, we allow  a central extension of the automorphism group.

The {\it fluctuation $\ddf$} of the Dirac operator $\mathcal{D}$ is given by a
finite collection $f$ of real numbers
$r_j$ and algebra automorphisms $\sigma _j\in{\rm Aut}(\mathcal{A})^e$ such
that
\bb
\ddf :=\sum_j r_j\,\mathbb{L}(\sigma _j) \, \mathcal{D} \, \mathbb{L}(\sigma_j)^{-1},\quad r_j\in\mathbb{R},\
\sigma _j\in{\rm Aut}(\mathcal{A})^e.
\eee
 We consider only fluctuations
with real coefficients since
$\ddf$ must remain selfadjoint.
The sub-matrix of the fluctuated Dirac operator $\ddf$ which is equivalent to
the mass matrix $\mathcal{M}$,  is often denoted by $\varphi $, the
`Higgs scalar', in physics literature. But one has to be careful, as will be shown
below explicitly. It may happen that the lifted automorphisms commute with
the initial Dirac operator and one finds $\ddf =\sum_i r_i \mathcal{D}$ for the finite
part of the spectral triple. This behaviour appeared for the first time in
the electro-strong model in \cite{class}, where the fermions couple vectorially
to all gauge groups and no Higgs field appears. In the model presented
below, the spectral triple can be decomposed into a direct sum consisting
of the standard model and two new particles. The initial Dirac operator
of the new particles commutes with the corresponding part of the lift and
thus does not participate in the Higgs mechanism.

An almost commutative geometry is the tensor product of a finite
noncommutative triple with an infinite, commutative spectral triple. By
Connes' reconstruction theorem \cite{grav,av} it is known that the latter comes
from a Riemannian spin manifold, which will be taken to be any
4-dimensional, compact manifold.  The spectral
action of this almost-commutative spectral triple is
defined to be the number of eigenvalues of the Dirac operator $\ddf$ up to a cut-off $
\Lambda$. 
Via the heat-kernel expansion one finds, after a long and laborious calculation \cite{cc,mc2},
a Yang-Mills-Higgs action combined with the Einstein-Hilbert action and a cosmological
constant:
\bb
S_{CC}[e,A_{L/R},\varphi] &=& \mbox{tr} \left[ h \left( \frac{\ddf^2}{\Lambda^2} \right) \right] 
\nonumber \\ \nonumber \\
&=& \int_M \left\{ \frac{2 \Lambda_c}{16 \pi G} - \frac{1}{16 \pi G} R + a (5R^2 - 8 R_{\mu\nu}
R^{\mu\nu} -7 R_{\mu\nu\lambda\tau}R^{\mu\nu\lambda\tau}) \right. \nonumber \\
&& + \sum_i \frac{1}{2 g_i^2} \mbox{tr}\ F^{\ast}_{i \mu \nu} F_i^{ \mu \nu} + \frac{1}{2} 
(D_\mu \varphi)^{\ast} D^\mu \varphi \nonumber \\
&& +\lambda \mbox{tr} (\varphi^{\ast} \varphi)^2 - \frac{1}{2} \mu^2 \mbox{tr} (\varphi^{\ast} \varphi)
\nonumber \\
&& + \left. \frac{1}{12} \mbox{tr} (\varphi^{\ast} \varphi) R \frac{}{} \right\} \; \rm{d}V + \mathcal{O}(\Lambda^{-2})
\label{CCaction}
\ee
where $h:\mathbb{R}_+ \rightarrow \mathbb{R}_+$ is a positive test function.
The coupling constants are functions of the  first moments $h_0$, $h_2$ and $h_4$ of
the test function:
\bb
&\Lambda_c \!&=\alpha_1 \frac{h_0}{h_2} \Lambda^2, \; G= \alpha_2 \frac{1}{h_2} \Lambda^{-2}, \;
a = \alpha_3 h_4 , \nonumber \\
&g_i^2\!& = \alpha_{4i} \frac{1}{h_4}, \; \lambda = \alpha_5 \frac{1}{h_4}, \; \mu^2 = 
\alpha_5 \frac{h_2}{h_4} \Lambda^2.
\label{CCcouplings}
\ee
The curvature terms $F_{\mu \nu}$ and the covariant derivative $D_\mu$ are in the standard form 
of Yang-Mills-Higgs theory.
The constants $\alpha_j$ depend in general on the special choice of the matrix
algebra and on the Hilbert space,
i.e. on the particle content. For details of the computation we refer to
\cite{cc,mc2}. 

This action is valid at the cut-off $\Lambda$ where it ties together the coupling constants $g_i$  of 
the gauge connections and the Higgs coupling $\lambda$ since they originate from the
same heat-kernel coefficient. 
For the standard model with three generations the calculation of the gauge couplings 
in (\ref{CCcouplings})
imposes at $\Lambda$ conditions on the $U(1)_Y$, $SU(2)_w$ and $SU(3)_c$
couplings $g_1$, $g_2$ and $g_3$ comparable to those of grand unified theories:
\bb
5 \, g_1^2 = 3 \, g_2^2 = 3 \, g_3^2 
\ee
But since the lift of the automorphisms produces extra free parameters through the
$U(1)$ central charges the first equality can always be modified by a different 
choice of the central-charge. Therefore only the gauge couplings of
noncommutative gauge groups underlie constraints from the spectral action.

In the same way as for the gauge couplings the spectral action also implies
constraints for the quartic Higgs coupling $\lambda$ and the Yukawa
couplings. The full set of constraints for the standard model reads \cite{c06,
mc2,thum}:
\bb 
 3 \, g_2^2=  3 \, g_3^2= 3 \,\frac{Y_2^2}{H} \,\frac{\lambda}{24}\,= \,\frac{3}{4}\,Y_2\,.
\label{4con}
\ee
Here $Y_2$ is the sum of all Yukawa couplings $g_f$ squared, $H$ is  the sum of all Yukawa couplings raised to the fourth power. Our  normalisations are: $m_f=\sqrt{2}\,(g_f/g_2)\,m_W,$ $(1/2)\,(\partial  \varphi)^2+(\lambda/24)\,\varphi^4$.

As we will see in the following, the grand unified constraint $g_2^2 = g_3^2$ at the
cut-off $\Lambda$ is a very special case. It is valid for  the standard model
but in general it will not hold. The model presented in this paper is one more
example for different constraints for $g_2$ and $g_3$ at the cut-off energy.
For possible extensions of the standard model within the framework of 
almost-commutative geometry, these constraints may limit the particle
content in a crucial way. 

\section{The spectral triple}

The model presented here is based on the spectral triple of the standard model
with four summands in the matrix algebra \cite{class}. In contrast to previous 
extensions of the standard model \cite{chris,theta} the algebra is not enlarged:
\bb
\mathcal{A}= \mathcal{A}_{SM}=\mathbb{H} \oplus \mathbb{C}  \oplus M_3(\mathbb{C}) \oplus
\mathbb{C} \ni (a,b,c,d).
\ee
Instead we enlarge the standard model by adding an a priori arbitrary number of 
generations 
of $SU(2)_w$ vector doublets. As we will see later, anomaly cancelation also requires
 vectorlike  hypercharges.
The representation of the algebra for these new particles
is:
\bb
\rho_L (d)=  d \,1_2,
\;
\rho_R (b)= \bar{b} \, 1_2, 
\;
\rho_L^c (a)=    a,
\;
\rho_R^c (a)=  a .
\ee
One sees immediately the vectorlike coupling to the quaternion sub-algebra 
in the antiparticle part of the representation. This results in a vectorlike coupling
to the $SU(2)_w$ subgroup of the standard model. Requiring the model to be
anomaly free will induce  the vectorlike  hypercharge  coupling. Note that 
these vector doublets do not satisfy all the physical requirements which had
been put forward in \cite{class} to classify almost-commutative geometries.
Notably the requirement of an unbroken colour group is not satisfied since
the $SU(2)_w$ subgroup acts as a colour for the vector doublets and is broken
by the  Higgs mechanism.  

The complete representation for the model is  the direct sum of the standard model 
representation and the representation for the vector doublets:
\bb
\rho = \rho_{SM} \oplus \rho_{vec} \ \  {\rm with} \  \ \rho_{vec}(a,b,d) = \rho_L (d) \oplus \rho_R (b) \oplus \rho_L^c (a) \oplus \rho_R^c (a)
\ee
The same holds for the Hilbert space, $\mathcal{H} = \mathcal{H}_{SM} \oplus \mathcal{H}_{new}$. For $N$ generations of vector doublets their Hilbert space is
\bb
\mathcal{H}_{vec} = ( \mathbb{C} \otimes \mathbb{C}^2 \otimes \mathbb{C}^N)_L
\oplus (\mathbb{C} \otimes \mathbb{C}^2 \otimes \mathbb{C}^N)_R \oplus \ {\rm antiparticles}.
\ee
The dimension of $\mathcal{H}_{vec}$ depends on the number of generations 
$N$ of vector doublets  and
reads dim$(\mathcal{H}_{vec})= 8\, N $.

We will denote the spinors of the vector doublets  $\psi_1$ and $\psi_2$ which 
are both hypercharge singlets.
\bb
\pp{\psi_{1} \\ \psi_{2}}_L \oplus \pp{\psi_{1} \\ \psi_{2}}_R  \oplus \pp{\psi_{1}^c \\ \psi_{2}^c}_L \oplus 
 \pp{\psi_{1}^c \\ \psi_{2}^c}_R \in \mathcal{H}_{vec}.
\ee
The Dirac operator 
contains the masses of the vector doublets and a CKM-like matrix which mixes  
the generations in the case of $N \geq 2$. For a first analysis of the model we will consider the CKM-like matrix to be 
the unity matrix. A nontrivial mixing matrix may be interesting when considering
leptogenesis-like processes to explain the particle-antiparticle asymmetrie in the 
universe.

The Dirac operator for one generation
of vector doublets is 
\bb
D_{vec} = \pp{ 0 & \mathcal{M}_{vec} & 0 & 0 \\ \mathcal{M}_{vec}^* &0&0&0 \\
0&0&0& \overline{\mathcal{M}}_{vec} \\ 0&0&\overline{\mathcal{M}}_{vec}^{\,*}&0}
\ \  {\rm with}  \ \ \mathcal{M}_{vec} = m_\psi 1_2. \ee

From the Krajewski diagram, figure \ref{kra1} in the appendix, 
it is straightforward to see that
all the axioms for the spectral triple are fulfilled. A second possibility to realise
vector doublets is depicted in the second Krajewski diagram, figure \ref{kra2}, 
in the  appendix.
This model exhibits essentially the same features as the one presented above,
only the sign of the hypercharges for the vector doublets is reversed.

\section{The gauge group, the lift and the constraints}

The automorphisms  that have be lifted coincides with the group of unitaries of the noncommutative part of the algebra \cite{farewell}:
\bb
\mathcal{U}^{nc}(\mathcal{A})= SU(2)_w \times U(3) \ni (v,w).
\ee
Defining $u:= \det (w) \in U(1)$, the particle part of the lift  decomposes into a 
left-handed part and a right-handed part.
\bb
\mathbb{L} ( v, u^{p_1},u^{p_2} w ,u^{p_3})=\mathbb{L}_{L} ( v, u^{p_1},u^{p_2}  w) \oplus   \mathbb{L}_{R} (v,u^{p_2}  w,u^{p_3})
\ee
with $p_i,q_i \in \mathbb{Z}$. We will impose here that the standard model
remains unchanged, i.e. that all the charges are the standard ones. From the
standard model part of the lift follows  then  $p_1=-p_3=-1/2$ and
$p_2=1/6 - 1/3$ through the requirement of anomaly cancellation. 
This reduces $U(3)$ to 
$U(1)_Y \times SU(3)_c$ in the correct representation. 

The exact form of the lift for the vector doublets is given by
\bb
\mathbb{L}_{vec} (v, u^{p_1}, u^{p_3}) =
{\rm diag}[ u^{p_1}  v, u^{-p_3}  v].
\ee
which is automatically anomaly free. We see now that the hypercharges of 
the vector doublets have been determined by fixing the hypercharges of the
standard model. Therefore the almost-commutative version of vector doublets
is far more constrained than vector doublets in the general Yang-Mills-Higgs 
setting where the
hypercharges are free parameters.

\begin{figure}
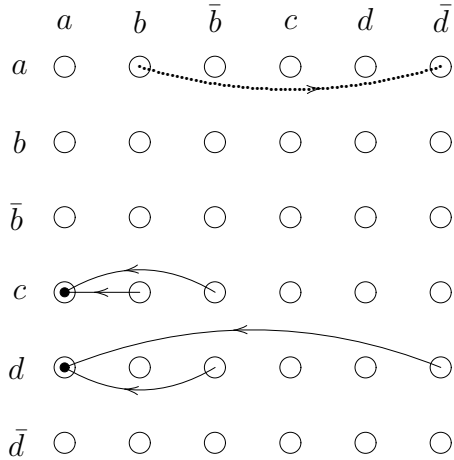

\begin{center}
\begin{tabular}{c}
\rxyh{0.7}{
,(5,-20)*\cir(0.3,0){}*\frm{*}
,(5,-25)*\cir(0.3,0){}*\frm{*}
,(5,-20);(10,-20)**\dir{-}?(.4)*\dir{<}
,(5,-20);(15,-20)**\crv{(10,-17)}?(.4)*\dir{<}
,(5,-25);(15,-25)**\crv{(10,-28)}?(.4)*\dir{<}
,(5,-25);(30,-25)**\crv{(17.5,-20)}?(.45)*\dir{<}
,(10,-5);(30,-5)**\crv{~*=<2pt>{.}(20,-8)}?(.6)*\dir{>}
}
\end{tabular}
\end{center}
\caption{ Krajewski diagram for the vector doublet model with reversed 
hypercharges.
}
\label{kra2}
\end{figure}

One sees immediately that the vector doublets have the same charge assignment
as the left-handed electron-neutrino doublet.
Therefore the electro-magnetic charge of the components of the vector doublets  are 
$Q_{el}= -e $ for $\psi_1$ and $Q_{el}= 0 $ for $\psi_2$. We will 
from now on call $\psi_1=:\psi^-$ and $\psi_2=:\psi^0$.
This charge assignment is summarised in table \ref{chargem}.
\begin{table}
\begin{center}
\begin{tabular}{|c|c|c|c|c|}
\hline
&I & $I_3$ & $Y_{vec}$ & $Q_{el}$  \\ 
\hline &&&&\\
$(\psi_{1})_{L,R} = \psi^-_{L,R}$ & $2$ & $-\frac{1}{2}$ & $-\frac{1}{2}$ & $-e$  \\
&&&& \\
\hline
&&&& \\
$(\psi_{2})_{L,R} = \psi^0_{L,R}$ & $2$ & $+\frac{1}{2}$ &$-\frac{1}{2}$ & $0$ \\
&&&& \\
\hline
\end{tabular}
\caption{Charge assignment for a negatively charged component}
\label{chargem}
\end{center}
\end{table}
For the model derived from the Krajewski diagram in figure \ref{kra2} one finds 
a hypercharge charge assignment with opposite signs and therefore opposite
electro-magnetic charges, see table \ref{chargep}.
\begin{table}
\begin{center}
\begin{tabular}{|c|c|c|c|c|}
\hline
&I & $I_3$ & $Y_{vec}$ & $Q_{el}$  \\ 
\hline &&&&\\
$(\psi_{1})_{L,R} = \psi^0_{L,R}$ & $2$ & $-\frac{1}{2}$ & $+\frac{1}{2}$ & $0$  \\
&&&& \\
\hline
&&&& \\
$(\psi_{2})_{L,R} = \psi^+_{L,R}$ & $2$ & $+\frac{1}{2}$ &$+\frac{1}{2}$ & $+e$ \\
&&&& \\
\hline
\end{tabular}
\caption{Charge assignment for a positively charged component}
\label{chargep}
\end{center}
\end{table}
We will from now on concentrate on the first case with $\psi^-$ and $\psi^0$.

Since the vector doublets couple have vectorlike couplings to the gauge group, 
the mass matrix $\mathcal{M}_{new}$ commutes with the lift $\mathbb{L}_{new}$
and it follows from the the inner fluctuations that the masses have no connection 
to the standard model Higgs field
\bb
\sum_j r_j\,\mathbb{L}_{L,vec} \, \mathcal{M}_{vec} \, \mathbb{L}^{-1}_{R,vec}
= \sum_j r_j \, \mathcal{M}_{vec}=:M_{vec} .
\ee
 Therefore $M_{vec}$ contains the gauge invariant masses of the 
 vector doublets where the real numbers $r_i$ are determined by the
 standard model part. 
 This phenomenon of gauge invariant masses in almost-commutative
 geometry appeared first in the case of the electro-strong model
 \cite{class}. It also appears in the $AC$ model \cite{chris} and in the 
 standard model with Majorana neutrinos \cite{mc2}. 
 
From the spectral action one obtains now immediately the Lagrangian for
the new particles,
\bb
\mathcal{L}_{vec} &=&
+  i \sum_{i=1..N} (\bar{\psi^-},\bar{\psi^0})^i_L D^\psi  \pp{\psi^- \\ \psi^0}^i_L 
+  i \sum_{i=1..N} (\bar{\psi^-},\bar{\psi^0})^i_R D^\psi  \pp{\psi^- \\ \psi^0}^i_R 
 \nonumber \\
&-&   \sum_{i=1..N} (\bar{\psi^-},\bar{\psi^0})^i_L M_{vec}^i  \pp{\psi^- \\ \psi^0}^i_R  +  \mbox{hermitian conjugate},
\ee
where the covariant derivatives are given by
\bb
D^{\psi} = \partial_\mu+ i g_1 \frac{Y_{vec}}{2} B_\mu +  i g_2 W^k_\mu \frac{\tau_k}{2} 
\ee
Here $g_1$ and $g_2$ are the standard model $U(1)_Y$ and $SU(2)_w$ gauge 
couplings with their corresponding generators. 

From the spectral action it is now straight forward to calculate the constraints
on the gauge couplings, the quartic Higgs coupling and the  Yukawa couplings
at the cut-off $\Lambda$.
The normalisation of the quartic
Higgs coupling is taken to be the same as for the standard model. Then the new
constraints read:
\bb
\left(3+ \frac{N}{2}\right) \, g_2^2= 3 \, g_3^2= 3 \,\frac{Y_2^2}{H} \,\frac{\lambda}{24}\,= \,\frac{3}{4}\,Y_2\,.
\label{4con}
\ee
$Y_2$ and $H$ include  the Yukawa couplings of the standard model
including a large Yukawa coupling for the $\tau$-neutrino
\cite{mc2}. We do not have any constraints on $g_1$ since the central 
charges that enter through the lift are free parameters.

This model has again the feature
that models beyond the standard
model in almost-commutative geometry will in general not exhibit 
the constraint $g_2=g_3$ from grand unified theories. A similar constraint
as in (\ref{4con}) already appeared in the model presented in \cite{theta}.

In the following we will neglect all standard model Yukawa couplings safe the top quark
coupling $g_t$ and the $\tau$-neutrino  Yukawa coupling $g_\nu$ which is adjusted 
to reproduce the correct top quark mass \cite{mc2}. We will not go into the
details of the seesaw mechanism in almost-commutative geometry but refer
to \cite{mc2} and \cite{jkss}. 
Under these  assumptions the relevant constraints on the couplings at the cut-off
$\Lambda$ read:
\bb
g_3^2 &=& \left(1+ \frac{N}{6}\right) g_2^2 \label{cutoff} \\
g_t^2 &=& \frac{4 + \frac{4 \, N }{6}}{3+ R^2} \, g_2^2 \ \ {\rm with}
\ \ R\,:= \, \frac{g_\nu}{g_t}  \label{gtop} \\
\lambda &=& 8 \, \left( 3 + \frac{N}{2} \right) 
\frac{3 +  R^4 }{(3 + R^2)^2 } \, g_2^2 \label{lambda}
\ee 
The ratio $R$ of  the
Yukawa coupling $g_\nu$ of the $\tau$-neutrino and  the top quark 
Yukawa coupling $g_t$ is fixed by the requirement that the renormalisation group
flow produces the measured pole mass of the top quark, $m_t=170.9\,\pm1.8$ GeV
\cite{data}.  

\section{Radiative corrections to the vector doublet masses}

The masses of the two components $\psi^-$ and $\psi^0$ of the vector doublet 
are a priori degenerate, $m_{\psi^-}=m_{\psi^0}=m_{\psi}$. 
But this degeneracy will split due to radiative corrections
for energies below the mass of the $Z$-boson $m_Z$. For a detailed
phenomenological discussion see \cite{wells}.

The calculations are well known and we will only give the result for the
mass splitting. Defining $r=\left(\frac{m_{\psi}}{m_{Z}}\right)^{2}$  one finds
for the mass difference 
 \bb
 \Delta m_{\psi}=\frac{\alpha}{2}M_{Z}f(r)
 \ \ {\rm with} \ \ f(r)= \int_{0}^{1}(2-x)\ln \left(1+\frac{x}{(1-x)^{2}r}\right),
 \ee
where the charged particle is heavier than its neutral partner, $m_{\psi^-}=
m_{\psi^0} + \Delta m_{\psi}$. 
Taking the limit $m_\psi \gg m_z$, i.e.  $r\gg1$,
 which will be interesting from the dark matter point of view, one finds the
 asymptotic mass difference   $\Delta m_\psi= \frac{1}{2} \alpha M_{Z} \simeq 355$
 MeV. 
It is also interesting to note that the lifetime of the charged particle is rather short
with 0.5 to 2 nanoseconds \cite{sher}. 

Since there are no terms in the Lagrangian coupling the vector bosons to standard
model particles we will consider the neutral particle as stable. It behaves essentially
like a neutrino and  its spin independent
cross section is of the same order, $\sigma_{si} \sim 10^{-39}$ cm$^2$. If these
particles are heavy enough $m_\psi > 10$ TeV they can escape direct detection
since current experiments are not sensitive for ultra massive dark matter particles.
Below $\sim 10$ TeV vector doublets can be excluded as a dark matter candidate
\cite{mdm}.
It has also been shown that neutrino-like particles with masses from $250$ TeV
to $550$ TeV may saturate the dark matter abundance of the universe \cite{griest}.
If the vector doublets are heavier than $550$ TeV they will over-close the universe.
We will therefore concentrate on the mass region between $10$ TeV and $550$ TeV. 

\section{The renormalisation group equations.}

We will now give the one-loop $\beta$-functions of the standard model   with three generations of standard model particles, $N$ generations of vector doublets.
They will serve to evolve the
constraints (\ref{4con}) from $E= \Lambda$  down to our energies $E=m_Z$. We set:
$ t:=\ln (E/m_Z),\qq \de g/\de t=:\beta _g,\qq \kappa :=(4\pi )^{-2}$.  
We will neglect the running of the gauge invariant masses of the 
vector doublets and treat them as free parameters. Furthermore 
all  threshold effects will be neglected.

The $\beta$-functions for the standard model with $N$ generations of vector
doublets are \cite{mv,jones}:
\bb 
\beta _{g_i}&=&\kappa b_ig_i^3,\qq 
b_i=
{\textstyle
\left( \frac{41}{6}  + \frac{2}{3}\,N,-\frac{19}{6}+\frac{2}{3}\,N,
-7  \right)  },
\\ \cr
\beta _t&=&\kappa
\left[ -\sum_i c_i^ug_i^2 +Y_2 +\,\frac{3}{2}\,g_t^2
\,\right] g_t,\\
\beta _\lambda &=&\kappa
\left[ \,\frac{9}{4}\,\left( g_1^4+2g_1^2g_2^2+3g_2^4\right)
-\left( 3g_1^2+9g_2^2\right) \lambda
+4Y_2\lambda -12H+4\lambda ^2\right] ,\ee
with
\bb c_i^t=\left( {\textstyle\frac{17}{12}},{\textstyle\frac{9}{4}} , 8 \right),
\
Y_2=3g_t^2, \
H=3g_t^4.
\ee
The Yukawa coupling of the $\tau$-neutrino can  be neglected in
the evolution of the renormalisation group equations, since the seesaw
mechanism renders it small compared to the top quark Yukawa coupling.

The gauge couplings decouple from the other equations and have  
identical evolutions in both energy domains:
\bb 
g_i(t)=g_{i0}/\sqrt{1-2\kappa b_ig_{i0}^2t}.
\ee
The initial conditions are taken from experiment \cite{data}:
\bb 
g_{10}= 0.3575,\qq
g_{20}=0.6514,\qq 
g_{30}=1.221.
\label{ggg}
\ee
Then the unification  scale $\Lambda $ is the solution of 
$\left(1+ \frac{N}{6} \right) \,g_2^2(\ln (\Lambda /m_Z))=
g_3^2(\ln  (\Lambda /m_Z))$,
\bb 
\Lambda = m_Z\exp\frac{g_{20}^{-2}-\left(1+ \frac{N}{6} \right) \, g_{30}^{-2}}{2\kappa (b_2-\left(1+ \frac{N}{6} \right) \,b_3)},
\ee
and  depends on  the number of generations
of vector doublets $N$. 

\section{The Higgs boson mass}

The aim is now to calculate the  mass of the Higgs boson,
$m_H$, fixing the quartic coupling $\lambda$ at the cut-off $\Lambda$
and then evolving it down to the pole mass with the renormalisation 
group equations.
We require that all couplings remain  perturbative and
we obtain the pole masses of the Higgs boson and  the top quark:
\bb 
m_H^2=\,\frac{4}{3}\,\frac{\lambda(m_H) }{g_2(m_Z)^2}\,m_W^2,\qq
m_t=\sqrt{2}\,\frac{g_t(m_t)}{g_2(m_t)}\,m_W.
\ee
As experimental input we have the initial conditions of the three 
standard model gauge couplings (\ref{ggg}) and the mass of the
top quark, $m_t=170.9\,\pm1.8$ GeV
\cite{data}.  As mentioned before the masses of the  vector doublets are 
taken to be between $10$ TeV and $550$ TeV. We will calculate the mass of 
the Higgs boson for these two extreme values
with the constraints (\ref{cutoff}) to (\ref{lambda}).

For the pure standard model we find a Higgs mass of $m_H = 
167.8^{+1.8}_{-1.7}$ GeV and a cut-off of $\Lambda = 1.1 \times
10^{17}$ GeV. This is in good agreement with
previous calculations \cite{mc2,jkss}.
We will now add subsequently one, two and three generations of 
vector doublets to the standard model. To simplify the analysis
we will assume the masses of the vector doublets to be equal
and the CKM mixing matrix to be trivial, i.e. the unity matrix. Nontrivial 
mixing between the generations may perhaps be interesting when
considering the particle-antiparticle asymmetry in cosmology.
Furthermore we will restrict ourselves to the two extrema of
the possible masses for the vector doublets: 10 TeV $\leq m_\psi \leq$
550 TeV.

For the case of one generation of vector doublets the Higgs mass
and the cut-off scale are summarised in  table \ref{one}.
\begin{table}
\begin{center}
\begin{tabular}{|c|c|c|}
\hline 1 generation & $m_{H}$ & $\Lambda$ \\
\hline $m_\psi=10$ TeV & $178,7^{+0,7}_{-0,7}$ GeV & $5.3\times10^{11}$  GeV \\
\hline $m_\psi=550$ TeV & $177,9^{+0,8}_{-0,7}$ GeV & $8.5\times10^{11}$ GeV \\
\hline
\end{tabular}
\caption{One generation of vector doublets}
\label{one}
\end{center}
\end{table}
Note that the cut-off scale is lowered considerably, by six orders of
magnitude with respect to the pure standard model. This phenomenon 
has two origins. On the one hand the running of the $SU(2)_w$ coupling
$g_2$ is diminished due to the presence of the vector doublets, while the
running of the 
$SU(3)_c$ coupling $g_3$ remains unchanged since the vector doublets 
are colour singlets.  Secondly the constraint (\ref{cutoff}) on $g_2$ and $g_3$ at the
cut-off gets modified. The effect of the vector doublets on the running of 
the couplings is rather small compared to their effect on the constraint (\ref{cutoff}).

For two generations of vector doublets  the  Higgs boson 
mass and cut-off scale are summarised in table \ref{two}.
\begin{table}
\begin{center}
\begin{tabular}{|c|c|c|}
\hline 2 generations & $m_{H}$ & $\Lambda$ \\
\hline $m_\psi=10$ TeV & $191,4^{+0,3}_{-0,2}$ GeV & $10^{9}$ GeV  \\
\hline $m_\psi=550$ TeV & $189,3^{+0,3}_{-0,2}$ GeV & $2.1\times10^{9}$ GeV  \\
\hline
\end{tabular}
\caption{Two generations of vector doublets}
\label{two}
\end{center}
\end{table}
One notes that the influence of the vector doublets on the Higgs mass is 
rather small. This should  compared to other models beyond the standard
model \cite{theta}, which can increase the Higgs mass by up to $\sim 160$ GeV
to $m_{Higgs} \sim 380$ GeV.
Table \ref{three} shows the Higgs boson masses and the  cut-off scales for three generations of vector doublets.
\begin{table}
\begin{center}
\begin{tabular}{|c|c|c|}
\hline 3 generations & $m_{H}$ & $\Lambda$  \\
\hline $m_\psi=10$ TeV & $204,9^{+0,3}_{-0,3}$ GeV & $2,8\times10^{7} $ GeV  \\
\hline $m_\psi=550$ TeV & $201,0^{+0,2}_{-0,1}$ GeV & $5,4 \times 10^{7}$ GeV  \\
\hline
\end{tabular}
\caption{Three generations of vector doublets}
\label{three}
\end{center}
\end{table}
To underline the general behaviour we also give a more extreme case with
five generations of vector doublets, see table \ref{five}.
\begin{table}
\begin{center}
\begin{tabular}{|c|c|c|}
\hline 5 generations & $m_{H}$ & $\Lambda$  \\
\hline $m_\psi=10$ TeV & $233,1^{+0,9}_{-0,9}$ GeV & $2,9\times10^{5} $ GeV  \\
\hline $m_\psi=550$ TeV & $224,4^{+0,5}_{-0,7}$ GeV & $8,0 \times 10^{5}$ GeV  \\
\hline
\end{tabular}
\caption{Five generations of vector doublets}
\label{five}
\end{center}
\end{table}
Here the cut-off scale has dropped down to the order of the vector doublet
masses. This is certainly a very interesting feature since it would give a 
natural explanation for  the  mass scale of the vector doublets. Furthermore
the Higgs mass is raised by $\sim 65$ GeV with respect to the pure standard
model. This allows the model to be clearly distinguished from the almost-commutative
standard  model by the LHC. The signature for this model would then be a very heavy 
Higgs boson and no further particles, since the masses of vector doublets should be
above the energy achieved by the LHC.

\section{Conclusions and outlook} 

We have presented a particle model based on an almost-commutative
geometry which contains the standard model as a sub-model. 
The spectral triple of the standard model is modified only slightly,
in the sense that the matrix algebra of the standard model stays 
unchanged and only an arbitrary number of $SU(2)_w$ vector doublets are 
added. 

These vector doublets are anomaly free, but their hypercharges
are constrained by the standard model hypercharges. This results
in an electro-magnetically  charged component of the doublet with one
electron charge and a neutral component.
Here again almost-commutative geometry is far more 
restrictive than general Yang-Mills-Higgs theory where in principle
any hypercharge for vector doublets is allowed and therefore
both components of  the doublets may be charged.
The masses of the vector doublets are gauge invariant, i.e. they
do not couple to the Higgs boson. Furthermore the new particles
are colour singlets with respect to the $SU(3)_c$ colour group
of the standard model.

The neutral particle in the doublet has a slightly lower mass than the charged
particle with a mass difference of $\Delta m_\psi \sim 350$ MeV.
This allows the charged particle to decay into its stable, neutral 
partner. The spin independent cross section of the neutral 
particle is of the same order of magnitude as a neutrino's
cross section, $\sigma_{si} \sim10^{-39}$ cm$^2$. If these
particles are sufficiently heavy they may be interesting dark 
matter candidates.

Considering masses for the vector doublets between 10 TeV
and 550 TeV one finds, when adding up to three generations
to the standard model, only a slight effect of $\sim35$ GeV
on the Higgs mass. In contrast the cut-off scale decreases 
considerably down to $\sim10^7$ GeV for three generations
of new particles. This low cut-off scale could explain in a very
natural way the scale of the gauge invariant masses of the
vector doublets.

Concluding one can certainly say that this model seems to be an interesting and 
promising extension of the standard model. Open issues are the direct
detectability of extremely heavy vector doublets by experiments such as 
EDELWEIS  and the effect of a nontrivial CKM-like mixing matrix.

\vskip1cm
\noindent
{\bf Acknowledgements:} The authors would like to thank T. Sch\"ucker, M.
Knecht and L. Lellouch for helpful comments and discussions. C. Stephan
gratefully acknowledges a fellowship of the Alexander von Humboldt-Stiftung.

\section*{Appendix: The Krajewski Diagram}

\begin{figure}
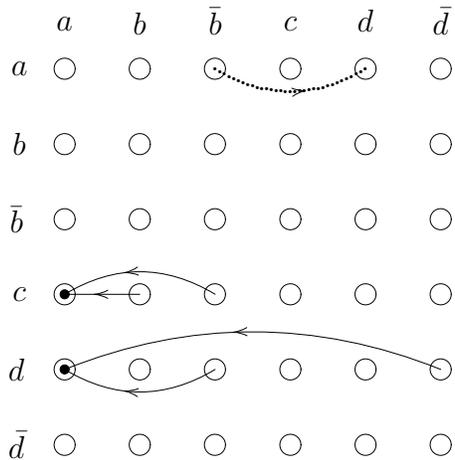

\begin{center}
\begin{tabular}{c}
\rxyh{0.7}{
,(5,-20)*\cir(0.3,0){}*\frm{*}
,(5,-25)*\cir(0.3,0){}*\frm{*}
,(5,-20);(10,-20)**\dir{-}?(.4)*\dir{<}
,(5,-20);(15,-20)**\crv{(10,-17)}?(.4)*\dir{<}
,(5,-25);(15,-25)**\crv{(10,-28)}?(.4)*\dir{<}
,(5,-25);(30,-25)**\crv{(17.5,-20)}?(.45)*\dir{<}
,(15,-5);(25,-5)**\crv{~*=<2pt>{.}(20,-8)}?(.6)*\dir{>}
}
\end{tabular}
\end{center}
\caption{ Krajewski diagram of the standard model with right-handed 
neutrino. The antiparticle part and the arrow representing  Majorana masses
has not been drawn. 
The  vector doublets are depicted by the dotted line.
}
\label{kra1}
\end{figure}

In this appendix we present the Krajewski diagrams which were
used to construct the model treated in this publication. 
Krajewski diagrams do for spectral triples  what the Dynkin and 
weight diagrams do for groups and  representations.
For an introduction into the formalism of Krajewski we refer to \cite{kps,class}.
The Krajewski diagram for the model presented in this paper 
is depicted in figure \ref{kra1}. It shows one generation of standard model
particles and one generation of vector doublets.


The arrows encoding the new particles are drawn on the $a$-line.
Note the similarity to the standard model quark
sector which sits on the $c$-line. 

The multiplicity matrix $\mu$ associated to the Krajewski diagram in figure \ref{kra1},
with three generations of standard model particles and $N$ generations
of vector doublets, can be directly read off to be
\bb
\mu = \pp{ 0&N&0&-N\\ 0&0&0&0 \\-3&6&0&0 \\ -3&3&0&3 }
\ee
The axiom of the Poincar\'e duality is fulfilled since
$\det (\mu - \mu^t ) = 36 \,  N^2 - 108 \, N + 81 \neq 0 \ {\rm for \ all} \ N \in \mathbb{N}$.
Only the right-handed neutrinos violate the  axiom of orientability, 
\cite{ko6}, which is also the case for the pure standard model. It is also
possible to reverse the arrow of the new particles, exchanging right-handed and
left-handed  vector doublets. But this does not change the general result.

\end{document}